\documentclass[aps,prd,twocolumn,groupedaddress,showpacs,nofootinbib,amssymb,balancelastpage,preprintnumbers]{revtex4}
\usepackage[dvipdfmx]{graphicx}
\usepackage{graphicx,bm,color}
\usepackage{amsmath}
\usepackage{amssymb}
\usepackage{amsfonts}
\usepackage{cases}
\usepackage{cancel}
\usepackage{hyperref}
\usepackage{ulem}
\usepackage{here}
\usepackage{subfigure}

\begin{document}
\preprint{RBRC-1325}

\title{
Nonperturbative Flavor Breaking in Topological Susceptibility 
at Chiral Crossover  
}  
\author{Mamiya Kawaguchi}\thanks{{\tt kawaguchi@fudan.edu.cn}} 
      \affiliation{ Department of Physics and Center for Field Theory and Particle Physics,
Fudan University, 220 Handan Road, 200433 Shanghai, China} 

\author{Shinya Matsuzaki}\thanks{{\tt synya@jlu.edu.cn}}
\affiliation{Center for Theoretical Physics and College of Physics, Jilin University, Changchun, 130012,
China}

\author{Akio Tomiya}\thanks{{\tt akio.tomiya@riken.jp}}
\affiliation{RIKEN BNL Research center, Brookhaven National Laboratory, Upton, NY, 11973, USA}

\begin{abstract}  

We demonstrate that the QCD topological susceptibility nonperturbatively 
gets a significant contribution signaled by 
flavor-nonuniversal quark condensates 
at around the pseudo-critical temperature of the
chiral crossover. 
It implies a remarkable flavor breaking in the axial anomaly as well as 
the QCD theta vacuum in high temperature QCD, which are almost flavor universal in the vacuum. 
A nontrivial flavor breaking is triggered by nonperturbative thermal loop corrections at around the chiral crossover, which is different from the trivial flavor violation just scaled by the quark mass ratio,  
observed at asymptotically high temperatures. 
This critical flavor violation cannot be 
dictated by 
the chiral perturbation theory with that lattice QCD usually compares, or the dilute instanton gas approximation based on that its 
astrophysical implications have 
conventionally been made. 
This would give an impact on 
the thermal history and the cosmological evolution of 
QCD axion including the estimate of the relic abundance as 
a dark matter candidate.  

\end{abstract} 
\maketitle

\section{Introduction}

The topological susceptibility 
is a crucial probe in studying the QCD $\theta$ vacuum structure and the axial anomaly. 
It is also important for QCD axion, 
which is postulated as an elegant solution 
to so-called the strong CP problem. 
In particular, when the QCD axion 
potential and mass are evaluated 
in the thermal history, 
the relic abundance as a cold dark matter today 
and the cosmological evolution of axion field in our Universe 
will be subject mainly to the temperature dependence of 
the topological susceptibility 
around the QCD phase transition epoch~\cite{Kim:2008hd}.  
 It would furthermore be a key quantity for 
hot QCD phenomena relevant to ongoing and designated heavy-ion collision experiments, 
such as physics 
induced by 
the topological charge fluctuation (i.e. susceptibility) closely tied with the presence of the QCD sphaleron
~\cite{Klinkhamer:1984di,McLerran:1990de,Arnold:1996dy,Huet:1996sh,Bodeker:1998hm}. 
Thus the QCD topological susceptibility has extensively been analyzed in multi-point of views with  
field theoretical, cosmological and astrophysical concerns.

The topological susceptibility $\chi_{\rm top}$ is defined as 
the curvature of the free energy of QCD with respect to 
the $\theta$ at the QCD vacuum with $\theta=0$. 
To our best knowledge, the key role for the $\chi_{\rm top}$ 
in real-life QCD is provided by an approximate chiral symmetry: 
with the approximate chiral symmetry for quarks, 
we can evaluate 
the $\chi_{\rm top}$ directly and generically from 
the QCD generating functional,  
in relation to 
the quark condensates $\langle \bar{q}q \rangle$  
and the current quark mass $m_q$~\footnote{ 
One may for instance refer to Appendix of the literature~\cite{Mao:2009sy}.}    \begin{align}
 \chi_{\rm top} 
 = \bar{m}^2(q) \sum_q \frac{\langle \bar{q} q \rangle}{m_q} 
 \,, \qquad 
 \frac{1}{\bar{m}(q)} \equiv \sum_q \frac{1}{m_q}
 \,,  \label{chi-top:formula}
\end{align} 
at the leading order of $m_q$ (omitting the higher-order  
pseudoscalar susceptibility terms). The relation in Eq.~(\ref{chi-top:formula}) 
 works even at finite temperature, 
because it is based on the Ward-Takahashi identity with the flavor singlet nature of the axial anomaly (as for the derivation of Eq.~(\ref{chi-top:formula}),   
see e.g. \cite{Kawaguchi:2020qvg}.).
This is the intriguing formula having the nonperturbative correlation 
in QCD  
between the axial anomaly along with the $\theta$ 
(dictated by the left hand side) and the chiral symmetry 
breaking (right hand side). 
The mediator for this correlation, 
the effective mass $\bar{m}(q)$ reflects   
the flavor-singlet nature of the axial anomaly in QCD~\cite{Baluni:1978rf}: 
the axial anomaly detected by the $\chi_{\rm top}$ goes away if either of quarks get massless.

In the QCD vacuum 
with the lightest three flavors 
($u,d,s$ having the mass well below the intrinsic QCD scale $\sim {\cal O}(1)$ GeV), 
the quark condensates are well degenerated ($\langle \bar{u}u \rangle 
\simeq \langle \bar{d}d \rangle \simeq \langle \bar{s}s \rangle \equiv \Sigma$) as observed in  
the recent lattice simulation, 
$\langle \bar{s}s \rangle/\langle \bar{l}l \rangle = 1.08 \pm 0.16$ 
($l=u,d$)~\cite{McNeile:2012xh}. 
In that case, the generic formula in Eq.(\ref{chi-top:formula}) 
can be reduced to the Leutwyler-Smiluga (LS) relation~\cite{Leutwyler:1992yt}: 
$\chi_{\rm top}|_{\rm LS}=\bar{m}(u,d,s) \Sigma$, 
which is derived at the leading order in 
the chiral perturbation theory (ChPT).  
This LS relation indeed works well in the QCD vacuum since 
the size of the three-flavor breaking is of ${\cal O}(\frac{m_l}{m_s} \frac{\langle \bar{s}s \rangle}{\langle \bar{l}l \rangle})={\cal O}(10^{-2})$, which supports 
the success of the ChPT also for this topological sector. The lattice QCD, with a physical pion mass  realized and the continuum limit taken,  
have also shown 
consistency with the ChPT prediction~\cite{Bonati:2015vqz,Borsanyi:2016ksw,Aoki:2019cca}.

The situation would get complicated at finite temperatures: at low temperatures enough, where 
the strange quark thermally decouples, the generic formula Eq.(\ref{chi-top:formula}) implies the $\chi_{\rm top}$ 
(normalized by the vacuum value)  
will keep almost constant reflecting the residual 
two-flavor ($u,d$) symmetry. 
This will be well described by the two-flavor ChPT. 
Given Eq.(\ref{chi-top:formula}), 
going to a fairly high temperature regime,  
where all the three quarks may decouple from gluons, so 
the $\chi_{\rm top}$ (normalized as such) could asymptotically get damped 
with quadratic power in temperature, just like the ideal quark gas picture. 

Highly nontrivial is, in particular, the response of the $\chi_{\rm top}$ 
at around the chiral phase transition: 
As has been observed in several analyses on hot lattice QCD~\cite{Cheng:2007jq,Nicola:2016jlj}, 
thermal loop effects 
would cause a partial restoration of the chiral symmetry (chiral crossover),  
where only the lightest $l$ quark condensates $\langle \bar{l} l \rangle$ 
drop, but the strange quark's $\langle \bar{s}s \rangle$ still keeps nonzero (or damps more slowly than the $\langle \bar{l}l \rangle$)\textcolor{black}{, as will be seen in the panel (b) of Fig.~\ref{condensates}.
}
This implies a nonperturbative flavor breaking in the quark condensates, hence would lead to a nonperturbative flavor breaking for the $\chi_{\rm top}$ 
at the hot QCD via the generic formula 
Eq.(\ref{chi-top:formula}), which surely cannot be captured by the ChPT, 
or the ideal quark gas picture, or even the semi-classical 
dilute instanton gas picture~\cite{Gross:1980br,Schafer:1996wv}.  

In fact, the recent lattice simulations~\cite{Bonati:2015vqz,Borsanyi:2016ksw} indicate  
a substantial deviation from the ChPT prediction  
on the temperature dependence 
of the $\chi_{\rm top}$ around and/or above  
the chiral-crossover transition point. 
This discrepancy should be understood 
by some nonperturbative analysis based on 
the generic formula Eq.(\ref{chi-top:formula}), 
though 
no definite implications to the 
$\chi_{\rm top}$ have been made in a sense of 
the quark-flavor symmetry at the crossover, even on the lattice QCD 
and chiral-effective model approaches.

In this Letter, we show  
that the QCD topological susceptibility indeed nonperturbatively 
gets a significant flavor violation signaled by 
a sizable strange-quark condensate-contribution 
at around the pseudo-critical temperature of the chiral crossover. 
It implies a nontrivial flavor breaking in the axial anomaly as well as the QCD theta vacuum at the chiral crossover.

We employ a nonperturbative analysis in a 
linear sigma model with the lightest three flavors, based on 
the Cornwall-Jackiw-Tomboulis (CJT) formalism~\cite{Cornwall:1974vz},
and compute thermal corrections to the effective potential 
coming from meson loops. 
The linear sigma model description allows us to go across 
the chiral phase boundary, in which 
the role of the chiral order parameter is played 
by sigma fields, as will be seen below. 

Some nonperturbative analyses on the 
QCD topological susceptibility at finite temperatures have so far 
been done based on chiral effective models~\cite{Fukushima:2001hr,Jiang:2012wm,Jiang:2015xqz}. 
However, no discussion on the correlation with 
quark condensates was made because their topological 
susceptibilities do not hold the flavor singlet form 
as in Eq.(\ref{chi-top:formula}) (see also Eq.(\ref{FSC}), for 
the flavor singlet condition), hence it seems 
to have been impossible to find the nonperturvative flavor breaking as addressed in the present Letter  
\footnote{
\textcolor{black}{
Another nonperturvative flavor breaking has been discussed based on Dyson–Schwinger analysis 
\cite{Horvatic:2019lok}.}
}
.



\section{Chiral Effective Model Description} 

We begin by introducing a linear sigma model with the lightest three flavors, which has recently been proposed by including a possible axial-anomaly induced-flavor breaking 
term~\cite{Kuroda:2019jzm}. 
We parameterize the scalar- and pseudoscalar-meson nonets by 
a $3\times 3$ matrix field $\Phi$ as 
$\Phi=(\sigma_a+i\pi_a)T_a$, where 
$\sigma_a$ are the scalar fields and $\pi_a$ are the pseudoscalar fields. $T_a=\lambda_a/2$ ($a=0,1,\cdots,8$) are the generators of $U(3)$ normalized by ${\rm tr}[T_aT_b]=1/2 \delta_{ab}$, where $\lambda_{a=1,\cdots,8}$ are the Gell-Mann matrices with $\lambda_0=\sqrt{2/3}\,{\bm 1}_{3\times 3}$.
Under the chiral $SU(3)_L\times SU(3)_R\times U(1)_A$ symmetry, 
$\Phi$ transforms as 
$\Phi \to g_A \cdot g_L \cdot \Phi \cdot g_R^\dagger$, 
where $g_{L,R}\in SU(3)_{L,R}$ and $g_A\in U(1)_A$. 
The three-flavor linear sigma model is thus written as \cite{Kuroda:2019jzm},
${\cal L} ={\rm tr}\left[ \partial_\mu \Phi\partial^\mu \Phi^\dagger \right]-V(\Phi) 
$
\,, 
with 
$ 
V(\Phi) = 
V_0+V_{\rm anom}+V_{\rm anom}+V_{\rm SB}+V_{\rm SB-anom}
\,.   
$
$V_0$ is an invariant part under the $SU(3)_L\times SU(3)_R\times U(1)_A$ symmetry, 
$V_0=\mu^2{\rm tr }[(\Phi^\dagger\Phi)]+\lambda_1{\rm tr }[(\Phi^\dagger\Phi)^2]+\lambda_2({\rm tr }[(\Phi^\dagger\Phi)])^2 . 
$
The chiral $SU(3)_L\times SU(3)_R$ invariant, but  
$U(1)_A$ anomalous part is incorporated in $V_{\rm anom}$: 
$ 
V_{\rm anom} 
=-B\left({\rm det}[\Phi]+{\rm det}[\Phi^\dagger]\right).
$

$V_{\rm SB}$ denotes the explicit chiral symmetry breaking term, which is originated from the current quark mass matrix ${\cal M} = {\rm diag}[m_u, m_d, m_s]$ in the 
underlying QCD Lagrangian. 
This ${\cal M}$ acts as a spurion field that transforms 
in the same way as the $\Phi$ does. 
At the leading order of expansion in $m_q$,  
the $V_{\rm SB}$ thus takes the form 
$V_{\rm SB}=-c{\rm tr}[{\cal M}\Phi^\dagger+{\cal M}^\dagger \Phi]. 
$
The parameter $c$ necessarily comes along with the current 
quark masses $({\cal M})$ in physical observables, which 
reflects the renormalization scale ambiguity in defining 
quark condensates.

$V_{\rm SB-anom}$ is the axial-anomaly induced-flavor breaking term. In the minimal flavor violation limit where single ${\cal M}$ is only allowed to be inserted, the $V_{\rm SB-anom}$ is cast into the form  
$ 
V_{\rm SB-anom}=-kc\left[\epsilon_{abc}\epsilon^{def}{\cal M}^a_d\Phi^b_e \Phi^c_f+{\rm h.c.}\right]. 
$
\textcolor{black}{
This non-standard axial-anomaly interaction induced by the flavor violation 
potentially makes $a_0(980)$ heavier than $K^*_0(700)$ because by construction, it supplies the strange quark mass only for $a_0(980)$.
Thus,
}
this $k$ coupling term plays a crucial role to 
realize the inverse mass hierarchy for 
scalar mesons below 1 GeV\textcolor{black}{: $m_{a_0(980)}\simeq m_{f_0(980)} >m_{K_o^*(700)}> m_{f_0(500)} $}, as has been shown in~\cite{Kuroda:2019jzm}. 
\textcolor{black}{
Moreover, intriguingly,
the axial anomaly-induced flavor breaking directly contributes to the quark condensates (see Eq.(\ref{ll-ss})), and then provides the nonperturbative flavor violation for the chiral crossover.}

In the QCD generating functional, the $\theta$-term can be rotated away by the $U(1)_A$ rotation with the rotation angle $\theta_q$ for 
quark fields. Then the $\theta$-dependence is fully transferred into 
the quark mass matrix, 
$\bar q_L {\cal M}q_R$ ($\bar q_R {\cal M}^\dagger q_L$) with
${\cal M}={\rm diag}[m_ue^{i\theta_u},m_de^{i\theta_d},m_se^{i\theta_s}]$. 
It is important to note here that 
the $\theta_q$ is constrained by the flavor singlet condition (for $\theta \ll 1$),
\begin{align} 
\theta_q=\frac{{\bar m}(u,d,s)}{m_q} \theta,
\label{FSC}
\end{align} 
so that the $U(1)_A$ anomaly keeps the flavor singlet nature~\cite{Baluni:1978rf}. 
We shall make a matching of this QCD generating functional with the one corresponding to the present linear sigma model. 
We are thus allowed to eliminate the $\theta$-dependence in the sigma field $\Phi$ ($\Phi^\dag$), which has been introduced as 
the interpolating field of quark bilinear $\bar{q}_R q_L$
($\bar{q}_L q_R$), 
so that 
the $\theta$-parameter should be entered in $V_{\rm SB}$ and $V_{\rm SB-anom}$ only via the quark mass matrix ${\cal M}$. 


We assume the isospin symmetry, $m_l=m_u=m_d\neq m_s$, so that
the vacuum expectation values are taken as 
$\langle \Phi \rangle={\rm diag }[\bar \Phi_1, \bar \Phi_2,\bar \Phi_3 ]$ 
with 
$\bar \Phi_1=\bar \Phi_2\neq \bar \Phi_3$.

All the introduced parameters 
$\mu^2, \lambda_{1,2}, B, c m_{l,s}, k$, and $\theta$ 
are taken to be real and positive.

\section{Nonperturbative Flavor Breaking in 
topological susceptibility}

To perform nonperturbative calculation of thermal loop contributions, 
we work on the CJT formalism~\cite{Cornwall:1974vz} 
by using the imaginary time formalism. 
The meson propagators contributing to thermal loops 
are then treated as full propagators determined 
by stationary conditions for the CJT effective potential 
($V_{\rm CJT}$), simultaneously 
with the vacuum expectation values $\bar{\Phi}_{1,3}$. 
Details \textcolor{black}{are reported in}~\cite{Kawaguchi:2020qvg}, instead, we shall here just show 
the result on the numerical analysis of the CJT formalism. 

\textcolor{black}{
The CJT potential is given by
\begin{eqnarray}
V_{\rm CJT}=
V_{\rm tree}(\bar \Phi_{1,3})+
V_{\rm 1 PI}(\bar \Phi_{1,3},S,P)+
V_{\rm 2 PI}(\bar \Phi_{1,3},S,P),
\end{eqnarray}
where $V_{\rm tree}(\bar \Phi_{1,3})$ is the tree-level potential and
$V_{\rm 1 (2) PI}(\bar \Phi_{1,3},S,P)$ arise from 
one (two)-particle irreducible diagrams, in which all meson loop lines are drawn by the dressed-full propagators, denoted as $S$ (for scalar mesons) and $P$ (pseudoscalar mesons). 
In the present study we work  
in the Hartree approximation, corresponding to the leading order approximation in the large $N$ expansion (only double-bubble diagrams survive for 
$V_{\rm 2 PI}(\bar \Phi_{1,3},S,P)$.
).
}
Since the CJT effective potential depends on the theta parameter $\theta$ through the quark mass matrix ${\cal M}$ 
as noted above, 
the topological susceptibility is straightforwardly computed 
as 
\begin{align} 
\chi_{\rm top}=
-\frac{\partial^2 V_{\rm CJT}}{\partial \theta^2}\Bigl |_{\theta=0}=
\left(\frac{2\langle \bar ll \rangle}{m_l}+\frac{\langle \bar ss \rangle}{m_s}
\right)\bar{m}^2(u,d,s)
\,, 
\label{chi_top_cjt}
\end{align} 
which is precisely in accordance with Eq.(\ref{chi-top:formula}). 
Terms suppressed by the higher order in $m_q$ have been omitted, as in Eq.(\ref{chi-top:formula}). 
The quark condensates are calculated also through the $V_{\rm CJT}$ as 
\begin{align} 
\langle \bar ll \rangle 
& =\frac{\partial V_{\rm CJT} }{\partial m_l}
= 
-2c(\bar\Phi_1+2k\bar\Phi_1\bar\Phi_3), 
\notag \\ 
\langle \bar ss \rangle
&=\frac{\partial V_{\rm CJT}}{\partial m_s} 
=-2c(\bar\Phi_3+2k\bar\Phi_1^2),  
\label{ll-ss}
\end{align} 
in which we will only take into account 
thermal correction parts~\footnote{Although the vacuum part has the ultraviolet divergences and  is subject to renormalization schemes, 
the thermal contribution to the loop integral is independent of the ultraviolet divergence and renormalization. 
Actually, the qualitative results obtained based on the CJT analysis at finite temperatures 
are fairly insensitive to renormalization schemes, as was studied in~\cite{Lenaghan:1999si,Lenaghan:2000ey,Roder:2003uz}.
Thus, as far as qualitaiive features deduced from nonperturbative 
corrections are concerned, 
we may be almost free from the ultraviolet sensitivity, allowing to  
just work on the thermal corrections from the meson loops. }.

Input parameters have been chosen by following the literature~\cite{Kuroda:2019jzm} as 
$ 
\mu^2 =1.02  
\times 10^4\,{\rm MeV}^2, 
\lambda_1= 11.8 
, 
\lambda_2=20.4 
, 
c m_l = 6.11 
\times 10^5\,{\rm MeV}^3, 
cm_s=198 
\times 10^5\,{\rm MeV}^3,  
B = 3.85 
\times 10^3\,{\rm MeV}, 
k=3.40 
\,{\rm GeV}^{-1}
$, which is a set of the best-fit values and 
well reproduces the scalar and pseudoscalar 
spectroscopy up to the mass scale of 1 GeV at the vacuum~\cite{Kuroda:2019jzm}.

As to the vacuum value of $\chi_{\rm top}$, 
the literature predicts 
$\chi_{\rm top}(T=0)=0.019(9)/{\rm fm}^4$~\cite{Bonati:2015vqz}, and 
$\chi_{\rm top}(T=0)=0.0245(24)_{\rm stat}(03)_{\rm flow}(12)_{\rm cont}/{\rm fm}^4$~\cite{Borsanyi:2016ksw}. 
For the latter the first error is statistical, the second error is systematic error and the third error comes from changing the upper limit of the lattice spacing range in the fit.  
Our prediction $\chi_{\rm top}(T=0)\simeq0.0263/{\rm fm}^4$ 
is in agreement with 
both those lattice observations with their systematic errors taken into account.

\begin{widetext} 

\begin{figure}[H]
\begin{center}
   \includegraphics[width=7.0cm]{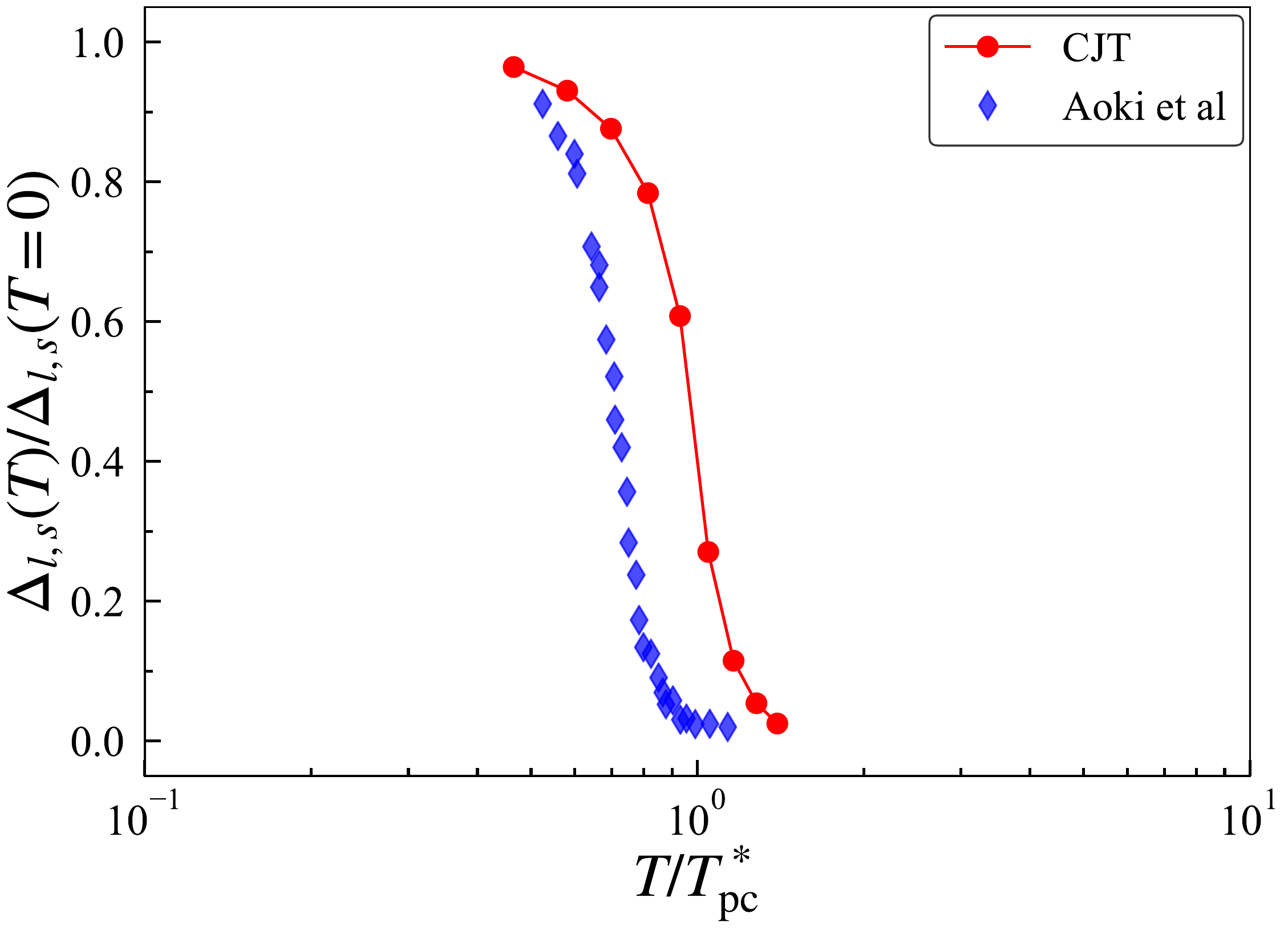}
    \subfigure{(a)}
   \includegraphics[width=7.0cm]{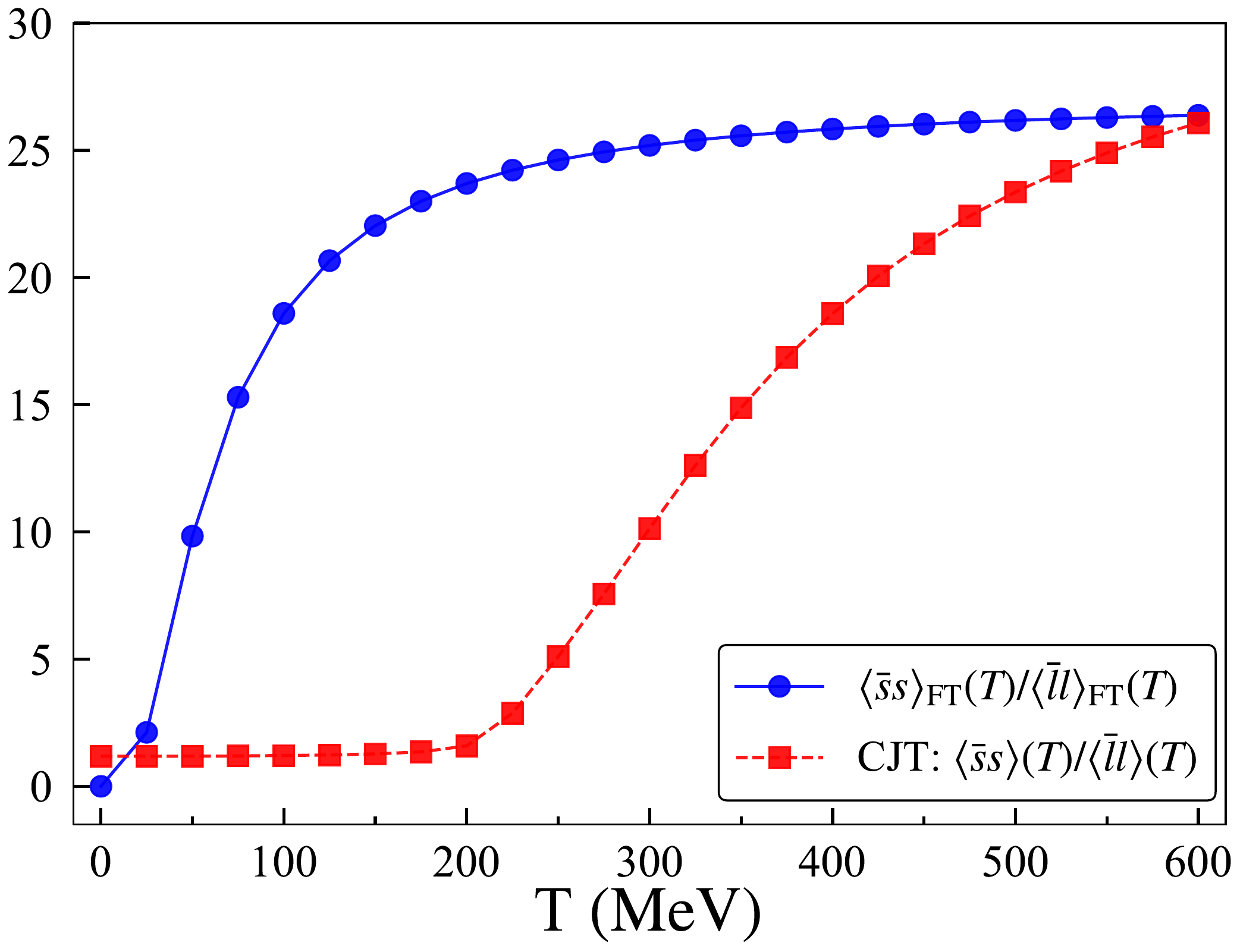}
    \subfigure{(b)}
  \end{center}
 \caption{The temperature dependence of the quark condensates.
 (a): comparison of the subtracted chiral condensate 
 $\Delta_{l,s}(T)=\langle \bar ll \rangle -\frac{2m_l}{m_s}\langle \bar ss \rangle$ with the lattice QCD data~\cite{Aoki:2009sc}; 
 (b): the ratio of quark condensates as a function of the temperature, $\langle \bar ss \rangle(T)/\langle \bar ll \rangle(T)$,
which implies the significance of the flavor breaking in the quark condensates as the temperature gets higher than the 
pseudo-critical temperature $T_{pc}^* \simeq 215$ MeV
while $\langle \bar ss \rangle(0)/\langle \bar ll \rangle(0)\simeq 1.18$~\cite{Kuroda:2019jzm}.
}  
 \label{condensates}
\end{figure}

\end{widetext}

By increasing temperature, we find 
the crossover ``phase transition" for the chiral symmetry in~\cite{Kawaguchi:2020qvg}.
Similar crossover phenomenon has also been 
observed in other three-flavor models based on the CJT formalism~ \cite{Lenaghan:2000ey,Roder:2003uz}. 
The pseudo-critical temperature $T_{pc}^*$ can simply 
be identified by 
$d^2 \langle \bar{l}l \rangle(T)/dT^2|_{T=T_{pc}^*}=0$, 
to be $\simeq 215$ MeV in the present analysis. 
This crossover phenomenon 
is in a qualitative sense consistent with 
the current result of the lattice QCD with $2+1$ flavors yielding 
the pseudo critical temperature $\simeq 155$ MeV~\cite{Aoki:2006we,Borsanyi:2011bn,Ding:2015ona,Ding:2020rtq}. 
Note that the definition of our pseudo-critical temperature $T_{pc}^*$ is different from the 
lattice QCD's and has been estimated to be 
larger than the lattice QCD value~\cite{Aoki:2006we,Borsanyi:2011bn,Ding:2015ona,Ding:2020rtq}: 
by construction of the present chiral effective 
model with single current quark mass matrix 
${\cal M}$ only included in operators, we cannot 
evaluate the chiral susceptibility, through which the pseudo-critical temperature in the lattice simulation is defined.   
However the temperature at which the maximum of the chiral susceptibility should reach is usually compatible with the inflection point of the chiral condensate in lattice QCD data, i.e. the point where 
$d^2 \langle \bar{l}l \rangle(T)/dT^2|_{T=T_{pc}}=0$. 
So, we may have quantitative comparison between them, 
to find about 30\% deviation for almost whole relevant 
temperature regime. See the panel (a) of Fig.~\ref{condensates}.

In the panel (b) of Fig.~\ref{condensates} we compare the $\langle \bar ss \rangle(T)/\langle \bar ll \rangle(T) $ with the one corresponding to the free theory of quarks, in which the quarks behave non-interacting free-particles. The ratio of free-quark condensates $\langle\bar ss \rangle_{\rm FT}(T)/\langle\bar l l \rangle_{\rm FT}(T)$ is perturbatively obtained from the thermal one-loop calculation. 
For the low temperature regions where $T<m_s \sim 100$ MeV, 
the free-theory (FT) quark-condensate ratio $\langle\bar ss \rangle_{\rm FT}(T)/\langle\bar l l \rangle_{\rm FT}(T)$ keeps the value below $m_s/m_l\simeq 27$, because of the naive Boltzmann suppression for the strange quark contribution. 
After passing $T\simeq m_s\sim 100$ MeV, the $\langle\bar ss \rangle_{\rm FT}(T)/\langle\bar l l \rangle_{\rm FT}(T)$ asymptotically approaches $m_s/m_l$, which merely reflects the trivial and overall flavor-breaking just by quark masses: $\langle \bar{q}q \rangle \sim m_q T^2$.
In contrast, in whole low-temperature regions  $T<T_{pc}^* \simeq 215$ MeV the CJT analysis exhibits a gigantic suppression 
for the quark-condensate ratio more than the Boltzmann suppression, to respect the flavor symmetry, i.e., the vacuum value $\sim 1$. 
This flavor symmetric behavior is consistent with the conventional three-flavor ChPT observation, 
in which the vectorial $SU(3)$ flavor symmetry cannot explicitly be violated at the leading order of the chiral expansion based on the nonlinear-sigma model setup, 
so that 
$\langle\bar ss \rangle_{\rm ChPT}(T)/\langle\bar l l \rangle_{\rm ChPT}(T)\simeq 1$ even including 
the next-to leading order corrections. 
As the temperature further increases, the CJT analysis nonperturbatively undergoes the chiral crossover around the pseudo-critical temperature $T_{pc}^* \simeq 215$ MeV, where the light quark condensate $\langle \bar ll \rangle(T)$ starts to drop more efficiently than the strange quark condensate $\langle \bar ss \rangle(T)$. 
Consequently, the quark-condensate ratio rapidly starts to grow from $T\simeq T_{pc}^*$.
This is a nontrivial flavor breaking, 
essentially different from the trivial FT flavor violation 
just by $m_s/m_l$. 
In the end, at around $T\simeq 600$ MeV, the quark-condensate ratio asymptotically merges with the FT yielding the trivial-flavor breaking value $m_s/m_l$. 
This would imply that the present linear sigma 
model would converge to an ideal-quark gas picture 
consistently with the asymptotic free nature of 
the underlying QCD. 
Thus we can conclude that what we call a nonperturbative flavor breaking is certainly a nonperturbative output in association with the characteristic chiral crossover phenomenon.

\begin{figure}[t] 
  \begin{center}
   \includegraphics[width=7.5cm]{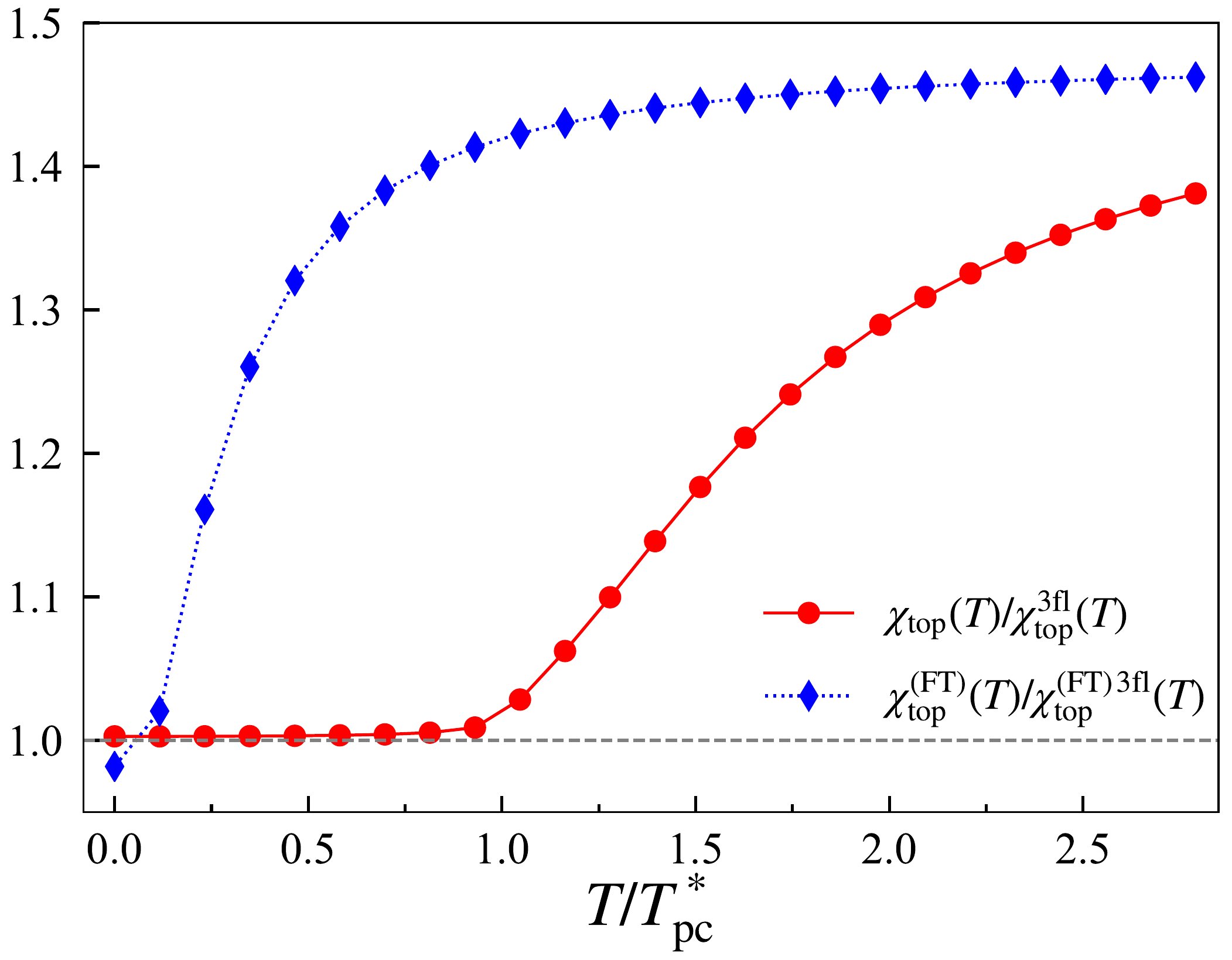}
  \end{center}   
 \caption{
The temperature dependence of the $\chi_{\rm top}(T)/\chi_{\rm top}^{3{\rm fl}}(T)$ compared with the free-quark theory, $\chi_{\rm top}^{({\rm FT})}(T)/\chi_{\rm top}^{({\rm FT})\,3{\rm fl}}(T)$.
}  
 \label{chitop}
\end{figure}

To extract the strange quark contribution in the topological susceptibility, in Fig.~\ref{chitop}
we show the temperature dependence of $\chi_{\rm top}(T)$ normalized to the one in the three-flavor universal limit, $\chi_{\rm top}^{3{\rm fl}}=\left(\frac{2}{m_l} + \frac{1}{m_s} \right)^{-1} \langle \bar ll \rangle(T)$.
We see that due to the sizable strange quark condensate as seen from the panel (b) of Fig.~\ref{condensates}, the topological susceptibility is rapidly enhanced from $T\simeq T_{pc}^*$. 
Eventually, after arriving at the high temperature regions where $\langle \bar ss\rangle(T)/\langle \bar ll\rangle(T)$ reaches the trivial-flavor breaking value $m_s/m_l$,  
the $\chi_{\rm top}(T)/\chi_{\rm top}^{3{\rm fl}}(T)$ merges with the quark-FT regime, to asymptotically converge to $\chi_{\rm top}(T)/\chi_{\rm top}^{3{\rm fl}}(T)\simeq 3/(2+m_l/m_s)\simeq1.5$. 
Thus, 
the topological susceptibility gets the nonperturbative flavor breaking at 
around the chiral crossover, which is manifestly different from 
the trivial-flavor breaking as seen in the quark-FT. 

\textcolor{black}{
Actually, it is obvious from the robust formula in Eq.(\ref{chi-top:formula})
that the strange quarks contribute to the $\chi_{\rm top}$, and, at high enough temperatures $T \gg T_{\rm pc}^*$, all flavors exhibit an equal importance as also deduced from the FT scaling in Fig.~\ref{condensates}. 
At low temperatures $T < T_{\rm pc}^*$, on the other hand, Eq.(\ref{chi-top:formula}) and the ChPT tells us that the contribution of the strange quarks is suppressed by a factor of approximately $m_s/m_{l} (\sim 27)$. 
What is nontrivial is that strange quarks start to become important in the $\chi_{\rm top}$ for $T \gtrsim T_{pc}^*$, as one can expect from Fig.~\ref{condensates} together with Eq.(\ref{chi-top:formula}), 
that has presently been made clarified in Fig.~\ref{chitop}, and we have dobbed 
the nonperturbative flavor violation in the $\chi_{\rm top}$. 
}

\begin{figure}[H]
\begin{center}
   \includegraphics[width=7.0cm]{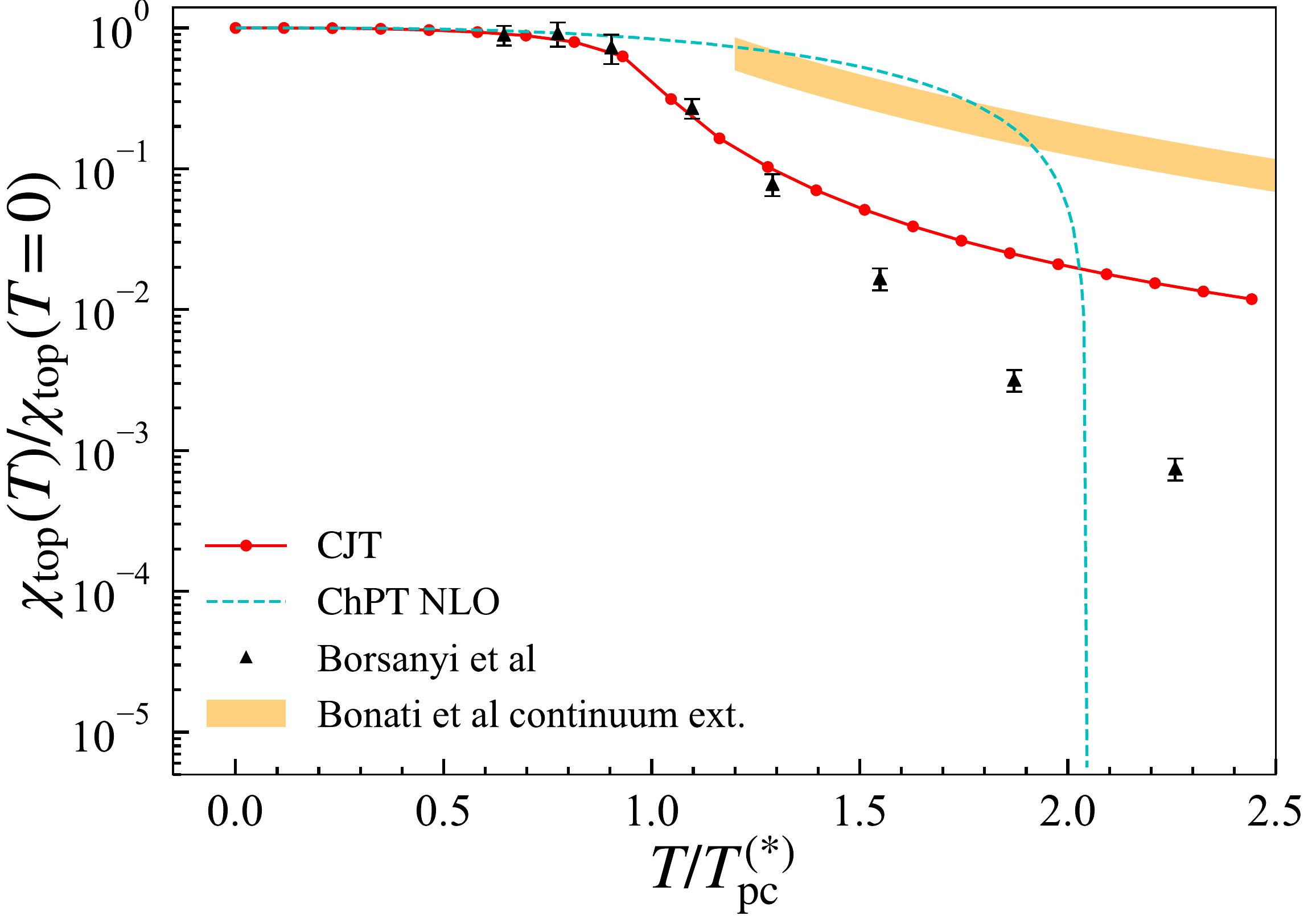}
    \subfigure{(a)}
   \includegraphics[width=7.0cm]{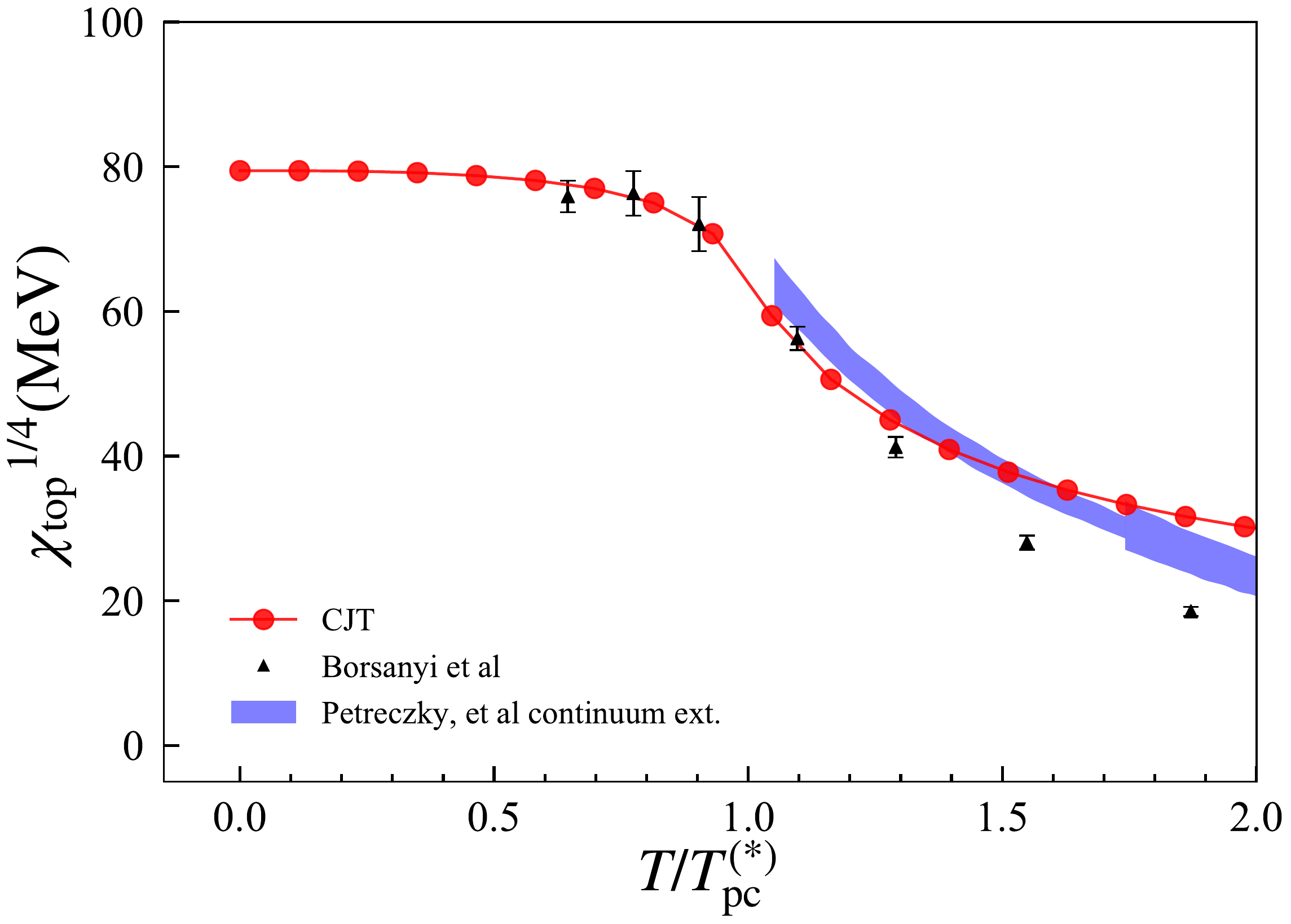}
    \subfigure{(b)}
  \end{center}
 \caption{
(a):
The comparison of the $\chi_{\rm top}(T)/\chi_{\rm top}(T=0)$ (labeled as ``CJT" in the plot) with the ChPT prediction up to the next-to-leading order (NLO)  (one-loop)~\cite{diCortona:2015ldu} and the lattice QCD data~\cite{Bonati:2015vqz,Borsanyi:2016ksw}. 
\textcolor{black}{We have taken $T_{\rm pc}^*=215$ MeV for the CJT result, and 
$T_{\rm pc}=155$ MeV for the lattice simulations and the ChPT prediction.}
The band corresponds to the continuum extrapolation of the lattice QCD data~\cite{Bonati:2015vqz}, which is estimated by the function
$\chi(a,T)/\chi(a,T=0)=D_0(1+D_1 a^2)(T/T_c)^{D_2}$ with 
$D_0=1.17$ $D_1=0$, $D_2=-2.71$ and $T_c=155$ MeV.
\textcolor{black}{
(b):
The unnormalized topological susceptibility, 
${\chi_{\rm top}}^{1/4}\,({\rm MeV})$, versus temperature, 
in comparison with 
the continuum extrapolated results 
of the recent lattice simulations.
The band corresponds to the continuum extrapolation of the lattice QCD data~\cite{Petreczky:2016vrs}.
}
}  
 \label{chitopvslat}  
\end{figure}

\textcolor{black}{The panel (a) of}
Fig.~\ref{chitopvslat} shows the $\chi_{\rm top}$ normalized 
to the vacuum value, 
in comparison with the ChPT prediction up to the next-leading order (one loop)~\cite{diCortona:2015ldu} and the recent lattice data with 2 + 1 (+1) flavors having a physical pion mass and the continuum limit being taken~\cite{Bonati:2015vqz,Borsanyi:2016ksw}. \textcolor{black}{
Since the linear sigma model 
including only mesons tends to systematically give a larger $T_{pc}^*$ 
than theories with quarks and mesons including lattice QCD,  
comparison with the lattice data has been made 
by normalizing $T$ by $T_{pc}^*$, so that such a 
systematic uncertainty is reduced. 
Note also that the linear-sigma model picture is thought to break down at high $T$, where degrees of freedom of quarks (and gluons) become important. Such a $T$ may be interpreted as $T \sim m_q^{\rm constituent} \sim 330$ MeV, where the latter denotes mass of constituent up and down quarks (given roughly by one-third of proton mass). 
In terms of the normalized $T$, it corresponds to $T/T_{pc}^* \sim 1.5$. Therefore, the present model prediction may be operative only   up until $T/T_{pc}^* \sim 1.5$, above which quarks and gluons 
would govern the system. As seen from Figs.1 and 2, however, 
it suffices for $T/T_{pc}^* \lesssim 1.5$ to explore the nonperturbative flavor violation in the $\chi_{\rm top}$, which 
starts to be eminent at around the chiral crossover $T/T_{pc}^* \gtrsim 1$. 
}

Remarkably, the predicted $T/T_{pc}^*$ dependence  
(denoted by ``CJT" in the figure) is actually slower-damping, and is overall 
consistent with both two lattice QCD data~\cite{Bonati:2015vqz,Borsanyi:2016ksw}~\footnote{
 \textcolor{black}{
Note that in Ref.~\cite{Bonati:2015vqz},
data have been taken only for two different lattice spacings. 
So,continuum extrapolation in Ref.~\cite{Bonati:2015vqz} may not reliably be  performed to give predictions. 
}
 }, 
which is not realized by the ChPT. 
This would manifest the importance of nonperturbative thermal contribution 
including the enhanced flavor breaking 
by $\langle \bar{s}s \rangle(T)/\langle \bar{l} l\rangle(T) \gg 1$ above $T_{pc}^*$, 
as depicted in Fig.~\ref{condensates}.

\textcolor{black}{
The CJT result favors the ChPT at $T/T_{pc}^* \lesssim 1$, which  implies that 
the chiral symmetry for light quarks might be essential in the low-temperature region, as discussed in the  literature~\cite{diCortona:2015ldu}~\footnote{
Note that the ChPT-governed domain is intact even if one includes  the next-to-next-to-leading order (NNLO) correction in the ChPT analysis~\cite{Nicola:2019ohb}. }. 
Beyond the ChPT-governed domain, above $T_{pc}$ the strange quark condensate 
would serve as an important source to develop the topological susceptibility, as the consequence of the nonperturbative flavor breaking.}

\textcolor{black}{The predicted curve also coincides with 
the dilute instanton gas approximation~\cite{Gross:1980br,Schafer:1996wv} for $T/T_{pc}^* \lesssim 1.5$, with which the lattice result in Ref.\cite{Borsanyi:2016ksw} is shown to be fully consistent.  
A substantial deviation starts in  
higher $T$, which might be because of lack of quarks (and gluons) in the present model. 
}


\textcolor{black}{In the panel (b) of Fig.~\ref{chitopvslat}, 
we also plot the temperature dependence of an unnormalized susceptibility  $\chi_{\rm top}^{1/4}({\rm MeV})$, where  
comparison with some recent lattice simulations with the extrapolation to the continuum limit is available for $T/T_{pc}^* \le 1.5$~\cite{Borsanyi:2016ksw, Petreczky:2016vrs}~\footnote{
\textcolor{black}{Another continuum extrapolated result in  Ref.~\cite{Bonati:2018blm} was provided for  $T/T_{pc}^* \gtrsim 3$}. 
}.
We see that at around the chiral crossover $T/T_{pc}^*\sim 1 - 1.5$ 
the CJT result is in quite good agreement with 
those continuum extrapolated data.  
Although 
qualitatively having agreement,  
for $T/T_{pc^*} \gtrsim 1.5$  
the CJT result tends to predict a somewhat larger $\chi_{\rm top}^{1/4}({\rm MeV})$. 
This discrepancy gets more larger as $T/T_{pc}^*$ 
gets larger and larger, which agrees with 
the expected validity of the present linear sigma model, 
as noted above. 
}

\textcolor{black}{
To reconcile the gap observed when $T/T_{pc^*} \gtrsim 1.5$, we may note that 
model parameters in the chiral effective model can 
actually have the intrinsic-temperature dependence, 
which could mimic a part of nondecoupling effects from 
integrating out quarks and/or gluons}. 
For instance, 
as noted in \cite{Pisarski:2019upw}, 
the instanton study predicts 
the parameter $B$ in the $U(1)_A$ anomalous part 
to have an intrinsic-temperature dependence, which is relevant to the QCD topological structure at high temperatures. Thus, such an 
intrinsic-temperature dependence might pull   
the CJT result down, to coincide with the lattice results in the continuum limit~\cite{Borsanyi:2016ksw, Petreczky:2016vrs,Bonati:2018blm}.


%

\section{Conclusion}

\textcolor{black}{In conclusion, we showed that the strange quark contribution to 
the topological susceptibility becomes eminent 
around the chiral crossover, 
which cannot be detected by the chiral perturbation theory 
and can clearly be distinguished 
from the trivial flavor breaking in high enough 
temperatures, just set by the quark mass weight.  
Our finding would be crucial to understand the hot QCD with the theta parameter, 
at around the chiral crossover
as well as the associated $U(1)_A$ anomaly.}  
In particular, the lattice QCD measurement of strange quark condensate at around and over the pseudo critical temperature would be crucial to check how the nonperturbative flavor breaking is critically operative 
in the topological susceptibility. 
This critical phenomenon should directly be observed 
in the future lattice QCD with appropriate accuracy. 

Another remark is that 
the topological susceptibility dumps slowly 
due to the nondecoupled nonperturbative strange quark condensate
which implies 
the delay of the effective $U(1)_A$ restoration.  
Thus the role of this nondecoupled strange quark condensate 
may account for the tension in the effective restoration of the $U(1)_A$ symmetry
currently observed on lattices with the two-flavor~\cite{Cossu:2013uua,Tomiya:2016jwr,Suzuki:2019vzy,Suzuki:2020rla}
and the 2+1 flavor~\cite{Bazavov:2012qja,Buchoff:2013nra,Ding:2019prx} near the chiral limit.  

Our proposal has been explicitized using some specific chiral effective model, especially including an axial 
anomaly-induced flavor breaking term. We have checked that 
even without this speciality, our finding is substantially unchanged, which \textcolor{black}{in details is reported in}~\cite{Kawaguchi:2020qvg}.  

The nonperturbative flavor breaking, especially, the significance of the strange quark condensate, in the 
topological susceptibility at around the pseudo critical temperature would give an  
impact on applications to QCD axion cosmological 
models. 
The epoch for QCD axion to start rolling 
and oscillate as well as the position in the potential 
is crucial to estimate the relic abundance of the axion as a dark matter today. This is subject to the temperature (or time) dependence 
of the $\chi_{\rm top}$ (which corresponds to the potential height) 
at around and/or above the chiral crossover boundary. 
As recently investigated in~\cite{Kim:2018knr}, 
below and above this crossover boundary, it has simply been 
assumed so far to apply the two-flavor ChPT and dilute instanton gas descriptions, 
separately, in evaluating the $\chi_{\rm top}$   
(with imposing a continuity condition between two domains through 
macroscopic thermodynamics quantities). 
Our present work would improve or refine this existing approach 
by including the strange quark contribution with use of the linear 
sigma model description, instead of the dilute instanton gas, 
at around and/or above the chiral crossover, and would provide 
a complementary evaluation of the QCD axion cosmology, with higher reliability. 
Detailed study deserves to another publication.


\section*{Acknowledgements} 

We are grateful to Massimo D’Elia, Xu-Guang Huang and Robert Pisarski 
for useful comments. 
This work was supported in part by the National Science Foundation of China (NSFC) under Grant No. 11747308 and 11975108 
and the Seeds Funding of Jilin University (S.M.).  
M.K.and A.T. thank for the hospitality of Center for Theoretical Physics and College of Physics, 
Jilin University where the present work has been partially done.  
The work of A.T. was supported by the RIKEN Special Postdoctoral Researcher program.



\end{document}